\documentclass[conference]{IEEEtran}

\usepackage{graphicx}
\usepackage{amsmath,amssymb,amsfonts}
\usepackage{multirow}
\usepackage{psfrag}
\usepackage{subfigure}
\usepackage{color}
\usepackage{soul}
\usepackage{xcolor}
\usepackage[hyphens]{url}
\usepackage[bookmarks=false]{hyperref}
\usepackage[hyphenbreaks]{breakurl}
\usepackage{lipsum}
\usepackage[ruled,vlined,linesnumbered]{algorithm2e}
\usepackage{mathrsfs}
\usepackage{pbox}

\DontPrintSemicolon

\makeatletter
\def\hlinewd#1{%
  \noalign{\ifnum0=`}\fi\hrule \@height #1 \futurelet
   \reserved@a\@xhline}
\makeatother

\graphicspath{{Figures/}}
\ifCLASSINFOpdf
\else
\fi

\begin{document}

\title{Characterizing Power Consumption of Dual-Frequency GNSS of Smartphone}

\author{Bikram~Karki and Myounggyu~Won\\
Department of Computer Science, University of Memphis, TN, United States\\
\{bkarki, mwon\}@memphis.edu
}

\markboth{Journal of \LaTeX\ Class Files,~Vol.~13, No.~9, September~2014}%
{Shell \MakeLowercase{\textit{et al.}}: Bare Demo of IEEEtran.cls for Journals}

\maketitle

\begin{abstract}
Location service is one of the most widely used features on a smartphone. More and more apps are built based on location services. As such, demand for accurate positioning is ever higher. Mobile brand Xiaomi has introduced Mi 8, the world's first smartphone equipped with a dual-frequency GNSS chipset which is claimed to provide up to decimeter-level positioning accuracy. Such unprecedentedly high location accuracy brought excitement to industry and academia for navigation research and development of emerging apps. On the other hand, there is a significant knowledge gap on the energy efficiency of smartphones equipped with a dual-frequency GNSS chipset. In this paper, we bridge this knowledge gap by performing an empirical study on power consumption of a dual-frequency GNSS phone. To the best our knowledge, this is the first experimental study that characterizes the power consumption of a smartphone equipped with a dual-frequency GNSS chipset and compares the energy efficiency with a single-frequency phone. We demonstrate that a smartphone with a dual-frequency GNSS chipset consumes 37\% more power on average outdoors, and 28\% more power indoors, in comparison with a singe-frequency GNSS phone.
\end{abstract}


\IEEEpeerreviewmaketitle

\section{Introduction}
\label{sec:introduction}

Location service is one of the most widely used features on a smartphone. According to the Ericson mobility report, more than 4.8 billion smartphones equipped with Global Navigation Satellite System (GNSS) chipsets are active worldwide in 2018~\cite{ericson}. More than 50\% of smartphone apps exploit location information~\cite{kaplan2005understanding} such as navigation, games, social media, dining service, to name a few. As the number of location service-based apps increases significantly, demand for higher positioning accuracy is ever higher. Unfortunately, provision of high localization accuracy has been limited and available only to professional and government use mostly due to the high cost and military security issue~\cite{robustelli2019assessment}.

However, a widespread use of smart devices is expanding the possibility of providing highly accurate positioning service to various apps on general user's smartphones and consequently resulted in the introduction of the world's first smartphone equipped with a dual-frequency GNSS chipset, Xiaomi Mi 8~\cite{mi8}. Xiaomi Mi 8 is equipped with a Broadcom's dual-frequency GNSS chipset (BCM47755)~\cite{broadcom} and is capable of receiving L1/E1 and L5/E5 signals from GNSS satellites~\cite{robustelli2019assessment}. More and more smartphones such as the Huawei Mate 20, Google Pixel 4, Lenovo Z6 Pro, LG G8, Samsung S10, and Sony Xperia 1 are being equipped with dual-frequency GNSS chipsets. With these phones that claim to provide decimeter-level positioning accuracy, a new dawn in location-based apps and navigation research is on the horizon.

One of the critical challenges for these dual-frequency phones is the energy efficiency. Although recent dual-frequency GNSS chipsets for these phones are equipped with advanced low-cost antennas that provide improved duty cycling to reduce the power consumption, dual-frequency GNSS chipsets require higher chipping rates and more processing, resulting in higher receiver power consumption to achieve high positioning accuracy. Numerous works have been performed on effectively utilizing dual-frequency GNSS phones focusing on achieving high positioning accuracy~\cite{wu2019precise}\cite{elmezayen2019precise}\cite{chen2019real}\cite{niu2019rtk}. However, how much power is consumed to achieve such high location accuracy is largely unexplored.

In this paper, we perform the first empirical study on power consumption of a smartphone equipped with a dual-frequency GNSS chipset. To the best of our knowledge, this is the first work that characterizes the power consumption of a dual-frequency GNSS phone in comparison with a smartphone equipped with a single-frequency GNSS chipset. For this study, two phones are selected, \emph{i.e.,} a currently available dual-frequency GNSS phone, Xiaomi Mi 8, and a single-frequency GNSS phone, Xiaomi Redmi Note 7, from the same vendor for fair comparative study. In particular, we measured the power consumption exclusively for updating positions performed by the GNSS modules of the phones by ruling out the energy consumption incurred by other hardware/software components of the phones. Experiments were performed both indoors and outdoors to understand the effects of different environments on the energy efficiency. We demonstrate that a phone with a dual-frequency GNSS chipset consumes 37\% more power on average for updating positions compared with its counterpart equipped with a single-frequency GNSS chipset outdoors, and 28\% more power in an indoor environment. Concretely, the contributions of this paper are summarized as follows.

\begin{itemize}
  \item We perform the first empirical study on the power consumption of a smartphone equipped with a dual-frequency GNSS chipset for both indoor and outdoor environments.
  \item We demonstrate that Xiaomi Mi 8 with a dual-frequency GNSS chipset consumes 37\% more power on average outdoors, and 28\% more power indoors, compared with Xiaomi Redmi Note 7 equipped with a single-frequency GNSS chipset.
  \item We present a useful reference on the energy efficiency of a dual-frequency GNSS phone to facilitate research on mobile computing and navigation that exploits those dual-frequency GNSS phones to achieve higher positioning accuracy.
\end{itemize}

This paper is organized as follows. In Section~\ref{sec:background} we review the background on GNSS concentrating on dual-frequency GNSS. We then explain the experimental settings and the methodology for measuring power consumption of both dual and single-frequency GNSS phones in Section~\ref{sec:exp_setup}. The results are analyzed in Section~\ref{sec:results}, and we conclude in Section~\ref{sec:conclusion}.

\section{Background}
\label{sec:background}

\subsection{Global Navigation Satellite System (GNSS)}
\label{sec:background_single}

GNSS is a system of satellites that provides time and location information anywhere on or near the Earth when an unblocked line of sight to four or more GNSS satellites is available~\cite{hofmann2007gnss}. There are two major GNSS systems that cover the entire world. Global Positioning System (GPS) is the most widely used system developed by US. It has at least 24 GNSS satellites. Globalnaya Navigatsionnaya Sputnikovaya Sistema (GLONASS) is a navigation system developed by Russian consisting of 31 GNSS satellites. There are two other systems with global coverage that are under development: BeiDou and Galileo. Beidou is a Chinese navigation system that has 22 satellites. While global coverage is not provided yet, it is already used in Asia-Pacific region. Galileo is the European navigation system consisting of 18 satellites. Full global coverage by Galileo is expected in 2020.

\begin{figure}[!htbp]
\centering
\includegraphics[width=.99\columnwidth]{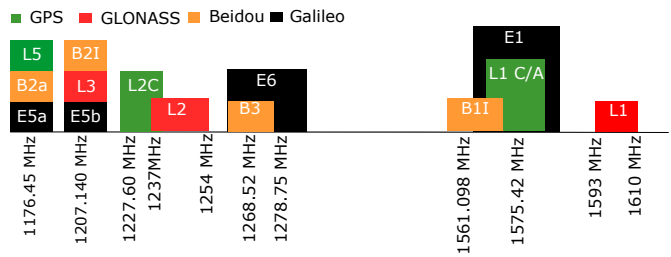}
\caption {GNSS frequencies for different navigation systems.}
\label{fig:lbandfrequency}
\end{figure}

GNSS satellites transmit radio signals over two or more frequencies in $L$ band, \emph{i.e.,} the operating frequency range of 1-2 GHz. Fig.~\ref{fig:lbandfrequency} shows the frequencies used by different GNSS systems. Radio signals transmitted from GNSS satellites carry ranging codes and navigation data which are used to calculate the coordinates of satellites and the distance between a satellite and a receiver. The binary phase shift keying (BPSK)~\cite{gardner1986bpsk} is used to modulate these signals.

A GNSS navigation message conveys various information such as the position and velocity of satellites, clock, satellite orbit shape, \emph{etc.} The navigation messages are transmitted at a slower rate than the ranging codes. For example, receiving a whole navigation message takes 30 seconds for GPS~\cite{wells1987guide}. The message consists of two types of data: Almanac and Ephemeris. The almanac data contain the coarse orbital parameters of all satellites and information about ionospheric delay corrections. Receivers use the almanac data to search satellites. Especially, receivers that support only a single GNSS frequency use the ionospheric delay data to correct the delay. To transmit the whole almanac data, 25 navigation messages are needed, and it takes about 12.5 minutes to complete the transmission. In contrast to the almanac data, the ephemeris data contain highly precise orbital parameters of satellites and clock correction information. The ephemeris data is used to calculate the positions of satellites precisely. Each satellite broadcasts its own ephemeris data every 30 seconds.

\begin{figure}[!htbp]
\centering
\includegraphics[width=.99\columnwidth]{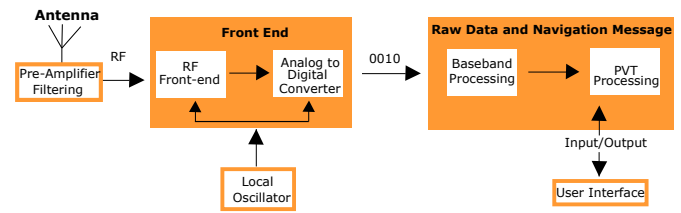}
\caption {Block diagram of a typical GNSS receiver.}
\label{fig:blockdiagram}
\end{figure}

A smartphone is equipped with a GNSS/navigation chipset to receive GNSS signals. It is kind of a blackbox that produces the user position, velocity and time (PVT) as well as information about tracked satellites. Fig.~\ref{fig:blockdiagram} shows a block diagram of a typical GNSS receiver. The GNSS antenna is used to capture GNSS signals in $L$ band. The RF front-end takes the RF signals as input from the antenna and performs down-conversion to reduce the cost. And then, the analog to digital converter (ADC) digitizes the signal. The baseband processing module performs several signal processing tasks to acquire and track the signals. The acquisition task determines satellites that are in view and can be tracked. The tracking stage is used to update dynamically the code delay and carrier frequency of the signal in order to track the signal correctly. The PVT processing block combines the information from the baseband processing block to derive a solution (\emph{e.g.,} PVT).

\subsection{Dual Frequency GNSS}
\label{sec:background_dual}

Dual-frequency GNSS receives two different radio signals at different frequencies from each satellite to provide more accurate positioning. Most of currently available devices utilize a single narrow band (L1/E1) with low sampling rates. Recently, the mass market introduce products that support dual wide band (upper $L$ band and partial lower $L$ band) with high sampling rates. However, due to the high cost, use of these dual-frequency GNSS devices has been limited to professional and governmental users. In 2017, Broadcom introduce the first low-cost dual-frequency GNSS chipset, BCM47755, specifically designed for smartphones~\cite{broadcom}. In 2018, u-blox launch their dual-frequency GNSS chipset, F9 chip~\cite{ubloc}, and STMicroelectronics introduce the Teseo receiver that supports L1/L2 or L1/L5 frequencies~\cite{stm}. Intel~\cite{intel} and Qualcomm~\cite{qualcomm} also start production of their dual-frequency GNSS chipsets in 2018.

With the growth of the mass market for dual-frequency GNSS chipsets, the first smartphone, Xiaomi Mi 8, equipped with a dual-frequency GNSS chipset, Broadcom's BCM47755, is introduced in 2018. This smartphone supports two frequencies (L1+L5) and is capable of tracking and processing GPS L1 C/A, GPS L5, GLONASS L1, Galileo E5a and QZSS L5, Galileo (GAL) E1, BeiDou (BDS) B1, GLONASS L1, and QZSS L1 signals.

These smartphones equipped with a dual-frequency GNSS chipset enjoy significant advantages. While enhanced positioning accuracy by directly estimating the ionosphere delay is the most significant benefit, the dual-frequency GNSS improves robustness against jamming and provides access to advanced satellite navigation technologies such as PPP~\cite{robustelli2019assessment} and RTK~\cite{dabove2019single}, which are currently available for only specialized receivers.

More precisely, errors from different sources influence the positioning accuracy such as imprecise information received from a satellite (\emph{e.g.,} on-board clock, ephemeris, \emph{etc.}), atmosphere effects (\emph{e.g.,} ionosphere and troposphere), receiver noise, and multipath effect~\cite{sanz2013gnss}. Among these various sources of errors, the ionosphere causes the biggest delay~\cite{kaplan2005understanding}. The ionosphere is a layer of the Earth's atmosphere extending from 60km up to 2,000km that contains a high concentration of free electrons and ions that can reflect radio waves. The impact of the layer is measured based on Total Electron Content (TEC) which is defined as the number of electrons in a tube of a 1m$^2$ cross section between two points, \emph{i.e.,} a receiver and a satellite. Thus, the contribution of the ionosphere can be written based on TEC as the following.

\begin{equation}
I_p = \frac{40.3 \cdot TEC}{f^2}.
\end{equation}

\noindent Here $f$ is the carrier frequency, and $40.3$ is the TEC parameter which depends on the location of receiver, the intensity of solar activity, and the hour of day. A dual-frequency GNSS chipset can eliminate the impact of the ionosphere based on a pseudorange $\rho_1$ calculated on frequency $f_1$ and a pseudorange $\rho_2$ calculated on frequency $f_2$ as follows. Details on calculating a pseudorange can be found in~\cite{groves2013principles}.

\begin{equation}
\rho^* = \frac{f_1^2\rho_1 - f_2^2\rho_2}{f_1^2 - f_2^2}.
\end{equation}

\noindent Here $\rho^*$ the pseudorange without the ionosphere effect.

There are challenges, however, to achieve enhanced accuracy. A dual-frequency GNSS impacts the design of the receiver, \emph{i.e.,} the antenna, RF front-end, and baseband processing blocks should be replicated to handle the additional frequency, leading to higher power consumption. Specifically, improved positioning accuracy is available with antennas with improved duty cycling for reducing power consumption. Utilization of dual bands requires higher chipping rates and more processing power resulting in degraded energy efficiency. This paper aims to characterize such extra power consumption for a dual-frequency GNSS phone by comparing that of a single-frequency GNSS phone.

\section{System Setup}
\label{sec:exp_setup}

\subsection{System Settings}
\label{subsec:device_under_test}

We select a currently available dual-frequency GNSS phone, Xiaomi Mi 8, and its single-frequency counterpart, Xiaomi Redmi Note 7 for this experimental study. Xiaomi Mi 8 is equipped with a dual-frequency GNSS chipset, Broadcom BCM47755, and Xiaomi Redmi Note 7 has a single-frequency GNSS integrated into its Qualcomm Snapdragon 660 processor.
Both phones have similar hardware specs as they are from the same vendor. They are equipped with Snapdragon processors, 16M color capacitive touchscreens, Adreno GPUs, and a similar set of sensors. We installed the same OS, Android Pie 9 (API level 28) on both phones. To cut off the effect of Assisted GPS~\cite{van2009gps} (\emph{i.e.,} a technique that utilizes the information from cell towers for faster position update), the sim cards of both phones were removed and Wi-Fi was disabled. An app was created that requests for a position update every second. All other apps including background processes were all disabled, and the brightness level of screen was kept to minimum.

\begin{figure}[!htbp]
\centering
\includegraphics[width=.6\columnwidth]{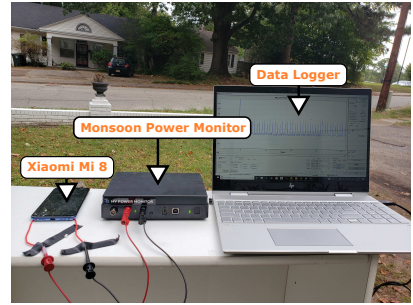}
\caption {Experimental setup.}
\label{fig:exp_setup}
\end{figure}

We used the Monsoon power monitor to measure power consumption~\cite{monsoon}. We deployed the system both indoors and outdoors. Fig.~\ref{fig:exp_setup} displays the system settings for outdoor deployment. The probes of the power monitor were connected to the battery terminals of smartphone, providing current to the phone. A laptop was connected to the Monsoon power monitor through USB to measure the current drawn and the voltage at a rate of 5KHz in real time.

\begin{figure}[!htbp]
\centering
\includegraphics[width=.8\columnwidth]{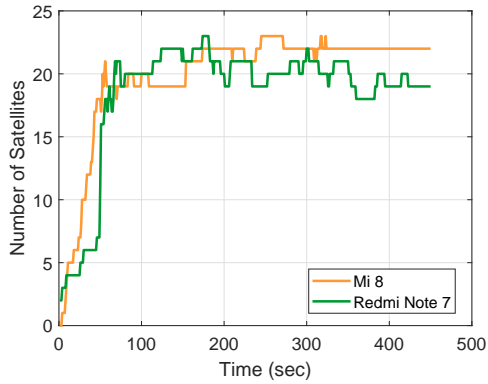}
\caption {The number of visible satellites outdoors.}
\label{fig:visible_sat}
\end{figure}

\begin{figure}[!htbp]
\centering
\includegraphics[width=.8\columnwidth]{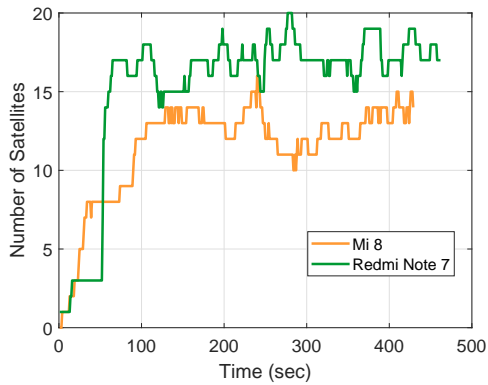}
\caption {The number of visible satellites indoors.}
\label{fig:visible_sat_indoor}
\end{figure}

We also confirm that both phones receive signals from a sufficient number of visible satellites. Figs.~\ref{fig:visible_sat} and~\ref{fig:visible_sat_indoor} display the number of visible satellites over time after the app is started outdoors and indoors, respectively. It shows that the number of visible satellites quickly increases when the app is started, and then the phones see about 18-22 satellites outdoors and about 14-18 satellites indoors. We observe that a sufficient number of satellites were available in both environments, although the number of visible satellites was smaller in the indoor environment.

\subsection{Methodology}
\label{subsec:methodology}

In this section, we present details on measuring the power consumption of the phones for updating positions. After GPS is switched on, it takes some time to complete the first position fix. This is called the time to first fix (TTFF)~\cite{paonni2010performance}. There are three different scenarios for TTFF. If GPS has been turned off for a long time and/or has moved a long distance, GPS does not have the almanac, ephemeris, time and position information. In this case, which is called the cold start, TTFF can be very large which may take several minutes. When only the ephemeris data is not available, which is called the warm start, TTFF can be significantly reduced as short as 30 seconds. If all the data are available, which is called the hot start, TTFF becomes minimal taking only 0.5 to 20 seconds.

\begin{figure}[!htbp]
\centering
\includegraphics[width=.8\columnwidth]{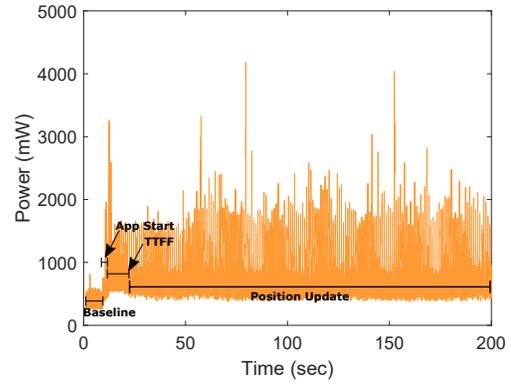}
\caption {Power consumption of Xiaomi Mi 8.}
\label{fig:mi8_power}
\end{figure}

\begin{figure}[!htbp]
\centering
\includegraphics[width=.8\columnwidth]{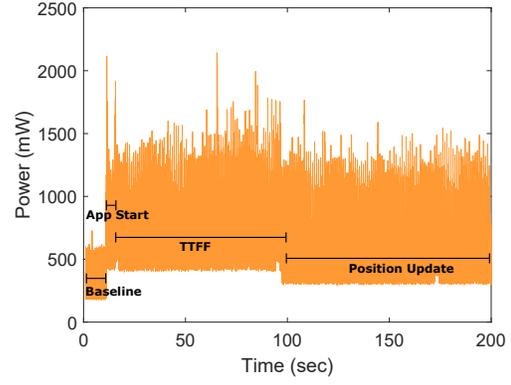}
\caption {Power consumption of Xiaomi Redmi Note 7.}
\label{fig:redmi_power}
\end{figure}

Figs.~\ref{fig:mi8_power} and~\ref{fig:redmi_power} show the power consumption of Mi 8 and Redmi Note 7. Both phones consume power independent of the GPS activity before the app is started. This is the baseline power consumption. Once the app is started, consumed power quickly increases for a short moment to load the app. And then, both phones consume relatively higher energy for TTFF compared to regular position update. The results show that TTFF for both phones were different. In fact, it is known that different GNSS chipsets have varying TTFF. Once the first position is fixed, both phones use power to update position. In this experiment, we focus on measuring power consumption for this regular position update, which accounts for the major part of power consumption for many apps based on location service.

\begin{figure}[!htbp]
\centering
\includegraphics[width=.8\columnwidth]{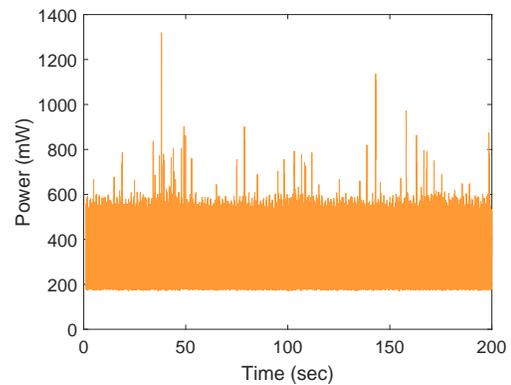}
\caption {Background power consumption of Xiaomi Redmi Note 7.}
\label{fig:background_red_mi}
\end{figure}

A challenge is to measure only the consumed power for updating positions, excluding other sources of power consumption such as background kernel processes, sensors, network modules, and screen. A tricky part is that these phones consume different amounts of power for these non-GPS activities. Fortunately, we found that the baseline power consumption of the two phones was relatively stable when we disable all background processes, disconnect network services such as Wi-Fi, and minimize the screen brightness to minimum as shown in Fig.~\ref{fig:background_red_mi}. Given the stable baseline power consumption, we simply subtract the baseline power consumption from measured power consumption and obtain the ``pure'' power consumption used for updating positions. More precisely, we set a one second interval within which a position update is performed, measure accumulated power consumption during this period, and then subtract it by the base line power consumption, obtaining consumed power for a single position update.

\section{Results}
\label{sec:results}


\subsection{Outdoor Experiments}

\begin{figure}[!htbp]
\centering
\includegraphics[width=.8\columnwidth]{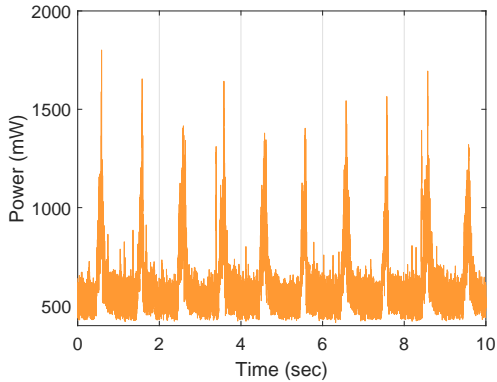}
\caption {Power consumption of Xiaomi Mi 8 for position update.}
\label{fig:outdoor_mi}
\end{figure}

\begin{figure}[!htbp]
\centering
\includegraphics[width=.8\columnwidth]{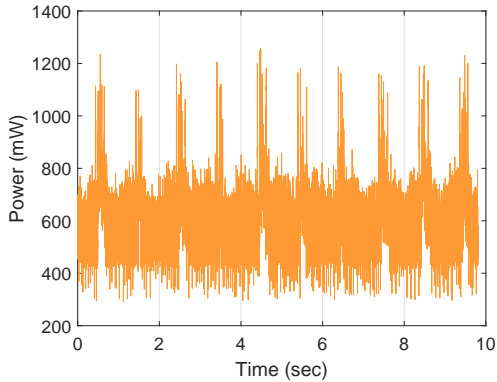}
\caption {Power consumption of Xiaomi Redmi Note 7 for position update.}
\label{fig:outdoor_note}
\end{figure}

\begin{figure}[!htbp]
\centering
\includegraphics[width=.8\columnwidth]{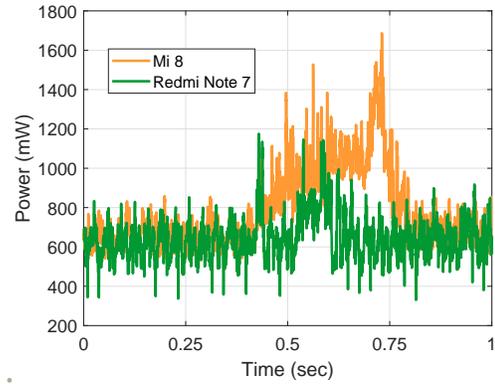}
\caption {Comparison of the power consumption of Xaiomi Mi 8 and Xiaomi Redmi Note 7 for a one second interval.}
\label{fig:comparison_close}
\end{figure}

Figs.~\ref{fig:outdoor_mi} and~\ref{fig:outdoor_note} display the power consumption of Mi 8 and Redmi Note 7 for updating positions, respectively. As shown, a peak is observed every second a request for a position update is sent to the phones. The results also show that Mi 8 with a dual-frequency GNSS chipset consumes more power than Redmi Note 7 with a single-frequency GNSS chipset. Fig.~\ref{fig:comparison_close} more clearly shows the difference in power consumption as we align the graphs exactly with the time when a request for position update is sent.

\begin{figure}[!htbp]
\centering
\includegraphics[width=.8\columnwidth]{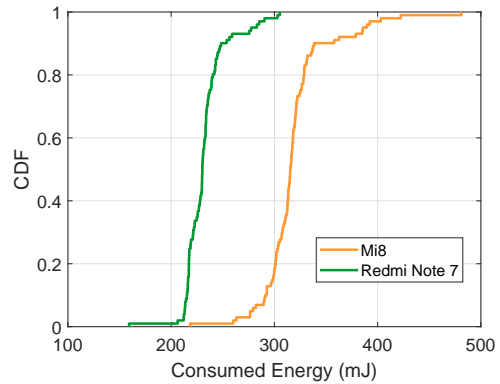}
\caption {Cumulative distribution function graph of the consumed energy of Xiaomi Mi 8 and Xiaomi Redmi Note 7 for updating positions.}
\label{fig:energy_for_pos_update_per_sec}
\end{figure}

We then calculate the ``pure'' power consumption for position update by subtracting with the baseline power consumption. We repeat a 5 minute measurement 5 times for each smartphone. Fig.~\ref{fig:energy_for_pos_update_per_sec} displays the cumulative distribution function (CDF) plots of power consumption for position update of both smartphones. The results demonstrate that the average power consumption for Mi 8 and Redmi Note 7 was 318mJ ($\pm$ 32mJ) and 232mJ ($\pm$ 20mJ), respectively, indicating that Mi 8 with the dual-frequency GNSS chipset consumes 37\% more power for position update in comparison with the single-frequency GNSS of Redmi Note 7. Considering that a typical phone battery has about 29,000 joules, without considering power consumption from any other hardware/software components, only the GPS for position update will deplete the battery after about 25 hours and 35 hours for Mi 8 and Redmi Note 7, respectively. It is interesting to note that a single-frequency GNSS phone would last 10 hours longer only due to the difference in the GNSS chipset.


\subsection{Indoor Experiments}

\begin{figure}[!htbp]
\centering
\includegraphics[width=.8\columnwidth]{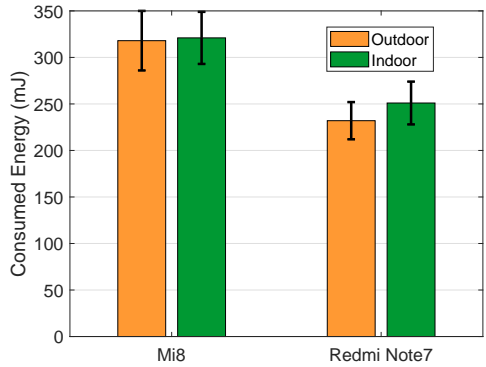}
\caption {Power consumption of Xiaomi Mi 8 and Xiaomi Redmi Note 7 in indoor and outdoor environments.}
\label{fig:indoor_vs_outdoor}
\end{figure}

\begin{figure}[!htbp]
\centering
\includegraphics[width=.8\columnwidth]{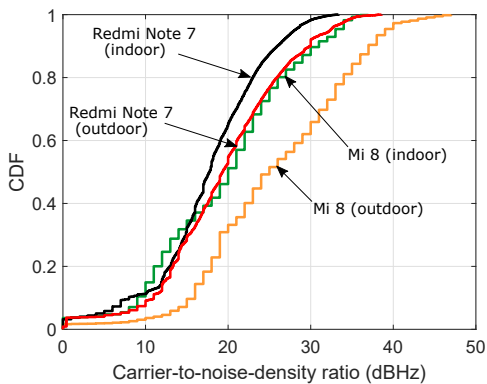}
\caption {Signal strength of GNSS signals (\emph{i.e.,} carrier-to-noise-density) for Mi 8 and Redmi Note 7 for indoor and outdoor environments.}
\label{fig:signal_strength}
\end{figure}

We measure the power consumption of Mi 8 and Redmi Note 7 used for updating positions in an indoor environment, \emph{i.e.,} inside an apartment. Fig.~\ref{fig:indoor_vs_outdoor} shows the results. As shown, the average power consumption of Mi 8 indoors and outdoors was 321mJ ($\pm$ 28) and 318mJ ($\pm$ 32), respectively. The average power consumption for Redmi Note 7 for indoors and outdoors was 251mJ ($\pm$ 23) and 232mJ ($\pm$ 20), respectively. The results indicate that more power is consumed in an indoor environment for both smartphones. Specifically, Mi 8 consumed 1.2\% more power, and Redmi Note 7 consumed 8\% more power in an indoor environment. Overall, in an indoor environment, Mi 8 consumed 28\% more power on average compared with Redmi Note 7. The reason for the higher power consumption in an indoor environment can be attributed to smaller signal strength of received GNSS signals in an indoor environment~\cite{lo2016greener} as shown in Fig.~\ref{fig:signal_strength}. The average signal strength for Mi 8 in an indoor environment decreased by 24\%, and that for Redmi Note 7 decreased by 10.5\%. Interestingly, it was found that the dual-frequency GNSS chipset of Mi 8 was less susceptible to weak signal strength compared with the single-frequency GNSS chipset of Redmi Note 7.

\section{Conclusion}
\label{sec:conclusion}

We presented the first empirical study on the power consumption of a dual-frequency GNSS smartphone. The measured power for a dual-frequency GNSS phone was compared with a single-frequency counterpart from the same vendor. We demonstrated that the dual-frequency phone consumed 37\% more power on average for position update compared with the single-frequency phone outdoors, and 28\% indoors. We expect that the results will be a useful reference for academia and industry in developing mobile applications exploiting location service based on dual-frequency GNSS.

\bibliographystyle{IEEEtran}
\bibliography{mybibfile}

\end{document}